\documentclass[journal]{IEEEtran}

\usepackage{multirow}
\usepackage[font=scriptsize,caption=false,labelsep=space]{subfig}
\usepackage{mathrsfs}
\usepackage{graphicx,float,epstopdf,amsmath}
\usepackage[]{algorithmicx}
\usepackage{algpseudocode,algorithm} 

\usepackage{color}
\usepackage[table]{xcolor}
\usepackage{amsmath}
\usepackage{amssymb}
\usepackage{amsthm}
\usepackage{graphicx}
\usepackage{epstopdf}
\usepackage{times}
\usepackage{textcomp,cite}
\usepackage{multicol}
\usepackage{cuted}
\usepackage{flushend}
\usepackage{balance}
\DeclareMathOperator*{\argmax}{argmax} 
\DeclareMathOperator*{\subjectto}{subject \hspace{3pt} to:}

\begin{document}
	
	\title{Hybrid Precoding for Multi-User Millimeter Wave Massive MIMO Systems: A Deep Learning Approach}
	
	\author{Ahmet M. Elbir and Anastasios Papazafeiropoulos\textit{, Senior Member, IEEE}
		\thanks{Copyright (c) 2015 IEEE. Personal use of this material is permitted. However, permission to use this material for any other purposes must be obtained from the IEEE by sending a request to pubs-permissions@ieee.org.
		
		A. M. Elbir is with the Department of Electrical and Electronics Engineering, Duzce University, Duzce, Turkey (e-mail: ahmetelbir@duzce.edu.tr).}
		\thanks{A. Papazafeiropoulos is with the Communications and Intelligent
			Systems Research Group, University of Hertfordshire, Hatfield AL10
			9AB, U.K., and also with SnT (http://www.securityandtrust.lu), University of Luxembourg, L-1855 Luxembourg City, Luxembourg (e-mail:	tapapazaf@gmail.com).}
		
	}
	
	%
	
	\maketitle
	
	\begin{abstract}
		In multi-user millimeter wave (mmWave) multiple-input-multiple-output (MIMO) systems, hybrid precoding is a crucial task to lower the complexity and cost while achieving a sufficient sum-rate.  Previous works on hybrid precoding were usually based on optimization or greedy approaches. These methods either provide higher complexity or have sub-optimum performance. Moreover, the performance of these methods mostly relies on the quality of the channel data. In this work, we propose a deep learning (DL) framework to improve the performance and provide less computation time as compared to conventional techniques.  In fact, we design a convolutional neural network  for MIMO (CNN-MIMO) that accepts as input  an imperfect channel matrix and gives the analog precoder and combiners at the output. The procedure includes two main stages. First, we develop an exhaustive search algorithm to {\color{black} select} the analog precoder and combiners {\color{black} from a predefined codebook}  maximizing the achievable sum-rate. Then, the selected precoder and combiners are used {\color{black} as output labels} in the training stage of CNN-MIMO where the input-output pairs are obtained. We evaluate the performance of the proposed method through numerous and extensive simulations and show that the proposed DL framework outperforms conventional techniques. Overall, CNN-MIMO provides a robust hybrid precoding scheme in the presence of imperfections regarding the channel matrix. On top of this, the proposed approach  exhibits less computation time with comparison to the optimization and codebook based approaches.

	\end{abstract}
	

	\section{Introduction}
	\label{sec:Introduciton}
	Millimeter wave (mmWave) communication systems provide a higher data rate and wider bandwidth at high frequencies   (in  the range of $30-300$ GHz) \cite{mimoOverview}. Reasonably, it has become a leading candidate to be realized in the fifth-generation (5G) wireless networks \cite{5GwhatWillItBe}. However, in mmWave bands, the propagation loss is higher as compared to conventional systems with lower frequencies \cite{mimoOverview,5GwhatWillItBe}. To overcome the high propagation path-loss and to provide beamforming power gain, massive numbers of antennas are used at both the transmitter and receiver sides by yielding a massive multiple-input-multiple-output (MIMO) structure  enhancing the signal-to-noise ratio (SNR) at the received signal \cite{mimoScalingUp}. 
	
	Signal processing in conventional systems with frequencies lower than 3GHz is performed digitally where both the amplitude and the phases are processed in the baseband. For this reason,  dedicated radio-frequency (RF) hardware for each antenna element is required \cite{mmWaveKeyElements}. Unfortunately, in the case of mmWave MIMO systems implemented with  a large number of antennas, digital processing is not cost-efficient since it brings high cost at the system hardware and significant complexity. 
	To reduce the cost and provide sufficient performance, hybrid precoding architectures are proposed where the signal is processed by both analog and digital precoders \cite{mimoHybridLeus1,mimoRHeath,Alkhateeb2015,hybridBFLowRes}. Especially, in the analog processing part of the hybrid systems, phase shifters with constant modulus are usually used. The role of phase shifters is the introduction of discrete phases to the transmitted/received  signal to steer the beam, and thus,  increase the gain \cite{hybridBFLowRes}.
	
	In recent years, several techniques have been proposed to design the hybrid precoding in mmWave MIMO systems. In particular, initial     works focused on the single-user scenario \cite{mimoRHeath}. In such a case, the user is assumed to be deployed with multiple antennas. While the single-user case constitutes the baseline for multi-user systems being of practical interest, the interference from other users should be taken into account when designing the  precoders \cite{hybridBFLowRes,mimoMU_simultaneousChannelEst,mimoMU_jointAS_HB,Alkhateeb2015}. In \cite{hybridBFLowRes}, the performance of low-resolution analog to digital converters (ADCs) are investigated when a single RF chain is used at mobile users. In \cite{mimoMU_simultaneousChannelEst}, simultaneous channel estimation is considered for multiple-user systems, while, in \cite{mimoMU_jointAS_HB}, antenna selection in mmWave MIMO is considered together with hybrid precoding estimation. The authors in \cite{Alkhateeb2015} also consider the multi-user scenario but the hybrid precoders are obtained by a greedy-like approach as in \cite{mimoRHeath} where a simultaneous orthogonal matching pursuit (SOMP) algorithm is proposed. It is worthwhile to mention that all of the above methods are based on the assumption of perfect channel state information and the availability of the array response sets,  namely, $\mathcal{F}$ and $\mathcal{W}$ for the precoder and combiner design, respectively. These sets are composed of the transmit and receive steering vectors with respect to the direction-of-arrival/departures (DOA/DODs) of the user locations. Taking into consideration that these array responses are directly related to the singular value matrix of the channel through a linear transformation, they become the best candidates for the precoder design problem \cite{mimoRHeath,mimoHybridLeus1,Alkhateeb2015}.

	As a class of machine learning techniques, DL has gained much interest recently for the solution of many challenging problems such as speech recognition, visual object recognition, and language processing \cite{deepLearningScience,deepLearning4SignalProcessing}. DL has several advantages such as low computational complexity when solving optimization-based or combinatorial search problems and the ability to extrapolate new features from a limited set of features contained in a training set \cite{deepLearningScience}. Very recently, a great deal of attention has been received for DL-based techniques regarding radar \cite{elbirIETRSN2019}, and fundamental communication theory topics \cite{deepLearningWC2,deepLearningWC3, mimoDLChannelEstimation, mimoDLChannelEstDOAEst,mimoDLHybrid,mimoDLDetection,mimoDL_reinforcementL,mimoDLOverAir,mimoDLOverAir2,mimoDLCSIFeedBack,Raj2018-at} such as channel estimation \cite{mimoDLChannelEstimation}, DOA estimation \cite{mimoDLChannelEstDOAEst}, and analog beam selection \cite{mimoDLHybrid}. Especially, in the physical layer of wireless communications, DL has been applied for signal detection \cite{mimoDLDetection}, channel estimation \cite{mimoDLOverAir,deepChannelEstBeamspace1,deepChannelEstBeamspace2}  and dynamic multi-channel access problems \cite{mimoDL_reinforcementL}.  In this direction, an end-to-end communication scenario is modeled in \cite{mimoDLOverAir} and \cite{mimoDLOverAir2} by using auto-encoders where single-input-single-output (SISO) systems are considered. The authors in \cite{mimoDLCSIFeedBack} have also used auto-encoders for the channel state information (CSI) feedback problem. Interestingly, \cite{Raj2018-at} studies the physical layer structures without channel models via DL. 
	
	An interesting topic concerns the  investigation of the hybrid precoding problem in the context of DL \cite{mimoDeepPrecoderDesign,deepMISO,mimoDLChannelModelBeamformingFacebook,elbirDL_COMML,elbirQuantizedCNN2019}.   Inspired from dense fully connected layers, deep multilayer perceptrons (MLPs) have been proposed in \cite{mimoDeepPrecoderDesign,mimoDLChannelModelBeamformingFacebook,deepMISO}. Specifically, in \cite{mimoDeepPrecoderDesign} and \cite{deepMISO},  MLP has been employed  only for the precoder design and just for the single-user scenario. In \cite{mimoDLChannelModelBeamformingFacebook}, an MLP architecture is considered for coordinated beam training where the perfect CSI is assumed to be known.	Moreover, in \cite{elbirDL_COMML}, a convolutional neural network (CNN)-based approach has been proposed for the joint precoder and combiner design problem but for the single-user setting again. Also, in \cite{elbirQuantizedCNN2019}, quantized and unquantized CNNs have been used for hybrid precoding in the case of a single-user MIMO system. {\color{black} The performance of DL-based approaches such as \cite{mimoDeepPrecoderDesign,mimoDLChannelModelBeamformingFacebook,deepMISO} strongly relies on the perfectness of the channel matrix whereas in \cite{elbirDL_COMML} and \cite{elbirQuantizedCNN2019}, robust DL approaches are proposed against the imperfections in the channel data but these works are developed only for the single-user scenario.}
	
	\subsection{Motivation}
	Although there are optimization-based approaches that directly estimate the precoders, they appear large computational complexity and local-minimum problems due to random initialization \cite{Yu2016-gn}. Also,  the design of hybrid precoders for the common  multi-user MIMO scenario, being of high practical importance, has not been considered in the context of DL. 
	Thus, driven by the advantages of DL such as its provided low computational complexity, we develop a method that can handle the hybrid precoding design in the case of multi-user MIMO transmission in the mmWave region when corrupted channel feedback data  is available. 
	
		\subsection{Contribution}
	{\color{black}In this paper, we propose a DL framework in terms of a CNN, which is  for mmWaves hybrid precoding design, henceforth  called CNN-MIMO.  In our DL framework, the channel matrix of users is selected as the input of CNN-MIMO, and the output labels are selected as the hybrid precoder weights. In the training stage, which is an offline process (please see Fig.~\ref{fig_Network}), we generate several channel realizations of multiple users and obtain the corresponding hybrid precoders via an exhaustive search algorithm. This process requires the knowledge of the feasible sets of array responses $\mathcal{F,W}$ which are not used in the prediction stage.  Once the network is trained, CNN-MIMO is used to predict the hybrid precoders by simply feeding the network with the channel matrix of users.
	} 
	{\color{black}The proposed DL framework provides a nonlinear mapping between the channel matrix and the hybrid beamformers. Hence, the proposed method achieves more robust performance than the competing algorithms since the deep network can handle the imperfections and the corruptions in the input channel data whereas the other algorithms do not have such capability. The proposed approach also has superior sum-rate performance due to the use of the ``best" hybrid beamformers which are obtained via an exhaustive search in the training process.} The main contributions of this work are as follows.
	\begin{itemize}
		\item A DL-based approach is proposed for the hybrid precoding in multi-user massive MIMO mmWave systems. We leverage DL to estimate the precoder and combiner weights so that CNN-MIMO is more robust against the deviations in the channel matrix. Hence, the proposed DL framework has superior performance with comparison to the conventional greedy and codebook based  techniques~\cite{mimoRHeath,Alkhateeb2015,hybridBFLowRes} whose performances strongly rely on the quality of the channel.  
		\item In most of the previous works such as \cite{mimoRHeath,Alkhateeb2015}, the codebooks formed by the feasible set of array responses $\mathcal{F}$ and $\mathcal{W}$ are assumed to be known. Then, the analog precoding design problem reduces to the selection of the best candidates in $\mathcal{F}$ and $\mathcal{W}$ to maximize the sum-rate. In this work, {\color{black} we only need $\mathcal{F}$ and $\mathcal{W}$ in the training stage to obtain the network labels and the proposed DL technique does not require such information in the prediction stage where} DL network itself obtains the analog precoder weights by learning the features hidden in the input data.
		\item To train the network, a very large training data (almost half a million samples) is generated. Hence, a robust performance against the imperfect channel case and the deviations in the channel data is achieved. 
		\item The proposed approach also enjoys less computation time for hybrid precoding design. While the conventional techniques require an optimization process or greedy searches, our CNN approach estimates the precoders by simply feeding the network with the corrupted channel matrix. 
	\end{itemize}


	\subsection{Notation} 
	Vectors and matrices are denoted by boldface lower and upper case symbols, respectively. In the case of a vector $\mathbf{a}$, $[\mathbf{a}]_{i}$ represents its $i$th element. For a matrix $\mathbf{A}$, $[\mathbf{A}]_{:,i}$ and $[\mathbf{A}]_{i,j}$ denote the $i$th column and the $(i,j)$th entry, respectively.  $\mathbf{I}_N$ is the identity matrix of size $N\times N$, $\mathbb{E}\{\cdot\}$ denotes the statistical expectation, and $\|\cdot\|_\mathrm{F}$ is the Frobenious norm.  The notation $(\cdot)^{\dagger}$ denotes the Moore-Penrose pseudo-inverse while  $\angle\{\cdot\}$ denotes the angle of a complex scalar/vector while the notation, expressing  a convolutional layer with $N$ filters of size $D\times D$, is given by $N$@$ D\times D$. For a complex scalar $a=e^{j\varphi}$ with continuous phase $\varphi$, $Q({a})= e^{j\varphi_B}$ denotes the quantization operator where $\varphi_B$ is the quantized angle in $[0,2\pi]$ sampled with $2^{B}$ points.

	\section{System Model}
	\label{sec:SystemModel}
\begin{figure}[t]
		\centering
		{\includegraphics[draft=false,scale=0.35]{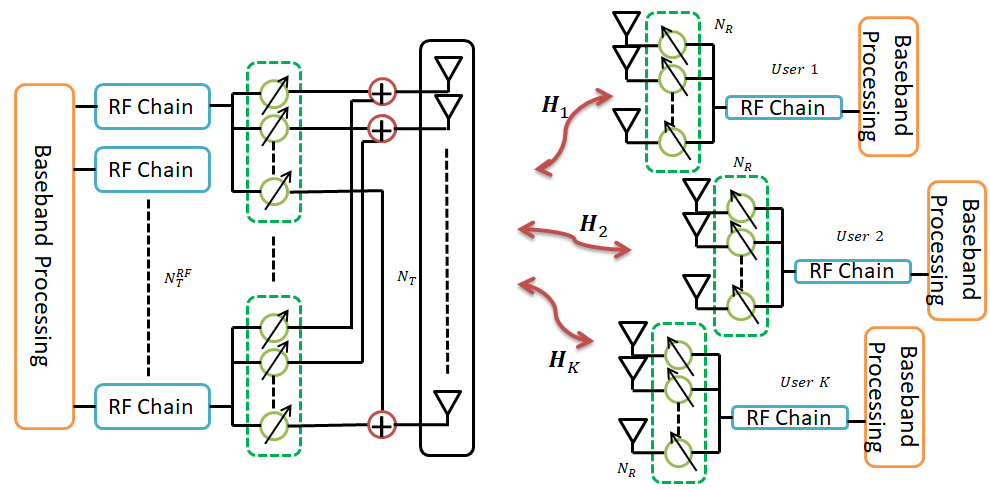} } 
		\caption{A multi-user MIMO system with hybrid  (analog and baseband) precoding on the BS and analog-only combining at $K$ users. }
		\label{fig_SystemModel}
	\end{figure}

	We consider a multi-user mmWave MIMO system as shown in Fig.~\ref{fig_SystemModel}. The base station (BS), serving $K$ users each of which has $N_\mathrm{R}$ antennas, is employed with  $N_\mathrm{T}$ antennas and $N_\mathrm{T}^{\mathrm{RF}}$ RF chains. By taking into consideration of cheaper hardware at each user, and subsequently, low power consumption, we assume that the BS communicates with each user via a single stream, i.e., $N_\mathrm{S}=1$~\cite{Alkhateeb2015}. Hence, only analog combining is applied at the receiver. Another assumption is that $ N_\mathrm{T}^{\mathrm{RF}} \ge K$, i.e., the maximum number of simultaneously served users cannot be greater than the number of BS RF chains. In the downlink, the BS applies baseband precoding $\mathbf{F}_{\mathrm{BB}} = [\mathbf{f}_{\mathrm{BB}_1},\mathbf{f}_{\mathrm{BB}_2},\dots,\mathbf{f}_{\mathrm{BB}_K}]\in \mathbb{C}^{N_\mathrm{T}^{\mathrm{RF}}\times K}$ to the transmit signal $\mathbf{s}=[s_1,s_2,\dots,s_K]^T\in \mathbb{C}^{K}$ obeying to  $\mathbb{E}\{\mathbf{s}\mathbf{s}^H\} = \frac{P}{K}\mathbf{I}_{K}$ by assuming equal power allocation among the users. Note that $P$ denotes the average power. The RF precoders $\mathbf{F}_{\mathrm{RF}}\in \mathbb{C}^{N_\mathrm{T}\times N_\mathrm{T}^{\mathrm{RF}}}$, which are constructed by phase shifters, are used to convey the signal to $N_\mathrm{T}$ transmit antennas. Also, given that $\mathbf{F}_{\mathrm{RF}}$ consists of analog phase shifters, we assume that the RF precoder has constant equal-norm elements, i.e., $|[\mathbf{F}_{\mathrm{RF}}]_{i,j}|^2 =1/N_\mathrm{T}$. In addition, we have the power constraint  $\|\mathbf{F}_{\mathrm{RF}}\mathbf{F}_{\mathrm{BB}} \|_\mathrm{F}^2= K$  that is enforced by the normalization of $\mathbf{F}_{\mathrm{BB}} $. Thus, the $N_\mathrm{T}\times 1$ transmitted signal is written as
	\begin{align}
	\mathbf{x} = \mathbf{F}_{\mathrm{RF}} \mathbf{F}_{\mathrm{BB}} \mathbf{s}.
	\end{align}

	We can write the received signal of the $k$th user for a narrowband block-fading channel  as \cite{mmWaveModel1}
	\begin{align}
	\label{def:ReceivedSignalAllAntennas}
	\tilde{\mathbf{y}}_k =\mathbf{H}_k \sum_{n=1}^{K}\mathbf{F}_{\mathrm{RF}}\mathbf{f}_{\mathrm{BB}_n}{s}_n + \mathbf{n}_k,
	\end{align}
	where $\mathbf{H}_k\in \mathbb{C}^{N_\mathrm{R}\times N_\mathrm{T}}$ is the channel matrix between the BS and the $k$th user   with $\|\mathbf{H}_k\|_\mathrm{F} = N_\mathrm{R}N_\mathrm{T}$. The vector $\mathbf{n}_k\in \mathbb{C}^{N_\mathrm{R}}$ denotes the complex additive white Gaussian noise (AWGN) with $\mathbf{n}_k \sim \mathcal{CN}(\mathbf{0}, \sigma^2 \mathbf{I}_{N_\mathrm{R}})$.

	Once the transmitted signal is received from the $k$th user, the received signal is processed by the combiner  $\mathbf{w}_{\mathrm{RF}_k}\in \mathbb{C}^{N_\mathrm{R}}$ as ${y}_k = \mathbf{w}_{\mathrm{RF}_k}^H\tilde{\mathbf{y}}_k$, i.e.,
	\begin{align}
	{y}_k 	= \mathbf{w}_{\mathrm{RF}_k}^H\mathbf{H}_k\sum_{n=1}^{K}\mathbf{F}_{\mathrm{RF}}\mathbf{f}_{\mathrm{BB}_n}{s}_n + \mathbf{w}_{\mathrm{RF}_k}^H\mathbf{n}_k,
	\end{align}
	where the RF combiners are constructed by means of phase shifters with the normalization constraint as $|[\mathbf{w}_{\mathrm{RF}_k}]_i|^2=1/N_\mathrm{R}$.
	
	\subsection{Channel Model}
	In mmWave transmission, the channel can be represented by a geometric model with limited scattering \cite{mimoChannelModel1,Raghavan06multi-antennacapacity,RappaportChannelModel}. Hence, we assume that the channel matrix  $\mathbf{H}_k$ includes the contributions of $L$ scattering paths. Considering a 2-D uniform planar array (UPA), the channel matrix corresponding to the $k$th user is given by
	\begin{align}
	\label{eq:ChannelModel}
	\mathbf{H}_k =  \gamma\sum_{l=1}^{L}\alpha_{l,k} g_\mathrm{R}(\Theta_\mathrm{R}^{\color{black}(l,k)}) g_\mathrm{T}(\Theta_\mathrm{T}^{\color{black}(l,k)})\mathbf{a}_\mathrm{R}(\Theta_\mathrm{R}^{(l,k)}) \mathbf{a}_\mathrm{T}^H(\Theta_\mathrm{T}^{(l,k)}), \nonumber
	\end{align} 
	where  $\Theta_\mathrm{R}^{(l,k)}=(\phi_\mathrm{R}^{(l,k)}, \theta_\mathrm{R}^{(l,k)})$ and $\Theta_\mathrm{T}^{(l,k)}=(\phi_\mathrm{T}^{(l,k)}, \theta_\mathrm{T}^{(l,k)})$  denote the angle of arrivals and departures, respectively. Note that the angular parameters $\phi$ and $\theta \in[0, 2\pi]$ correspond to the azimuth and the elevation angles, respectively. 
	The scalar $\gamma = \sqrt{ N_\mathrm{T} N_{\mathrm{R}}/L}$ is the normalization factor and $\alpha_{l,k}$ is the complex channel gain associated with the $k$th user and $l$th path $l = 1,\dots, L$. Also,  $g_\mathrm{R}(\Theta_\mathrm{R}^{(l,k)})$ and $ g_\mathrm{T}(\Theta_\mathrm{T}^{(l,k)})$ are the antenna element gains for the antennas in the arrays while $\mathbf{a}_\mathrm{R}(\Theta_\mathrm{R}^{(l,k)})$ and $\mathbf{a}_\mathrm{T}(\Theta_\mathrm{T}^{(l,k)})$  are the $N_\mathrm{R} \times 1$ and $N_\mathrm{T}\times 1$ steering vectors representing the array responses at the $k$th user and the BS, respectively. The $n$th element of the steering vector $\mathbf{a}_\mathrm{R}(\Theta_\mathrm{R}^{(l,k)})$ is given as
	\begin{align}
	[\mathbf{a}_\mathrm{R}(\Theta_\mathrm{R}^{(l,k)})]_n = \exp\left\{ -\frac{2\pi}{\lambda}\mathbf{p}_n^T \mathbf{r}(\Theta_\mathrm{R} ^{(l,k)}) \right\},
	\end{align}
	where $\lambda$ is the wavelength, $\mathbf{p}_n =[x_n,y_n,z_n]^T$ is the position of the $n$th antenna in the Cartesian coordinate system. Regarding the direction vector, it is given by
	\begin{align}
	\mathbf{r}(\Theta_\mathrm{R}^{(l,k)}) =& [\sin(\phi_\mathrm{R}^{(l,k)})\cos(\theta_\mathrm{R}^{(l,k)}), \nonumber \\
	&\sin(\phi_\mathrm{R}^{(l,k)})\sin(\theta_\mathrm{R}^{(l,k)}),\cos(\theta_\mathrm{R}^{(l,k)})]^T.
	\end{align}
	
	In a similar way, the transmitter side steering vector $\mathbf{a}_\mathrm{T}(\Theta_\mathrm{T}^{(l,k)})$ can also be defined as for $\mathbf{a}_\mathrm{R}(\Theta_\mathrm{R}^{(l,k)})$.
	
	By assuming that  Gaussian symbols are transmitted through the mmWave channel under study,  the achievable rate for the $k$th user is written as \cite{mimoHybridLeus1,Alkhateeb2015} 
	\begin{align}
	\label{eq:Rate}
	R_k  = \log_2 \bigg| 1 + 
	\frac{ \frac{P}{K} |  \mathbf{w}_{\mathrm{RF}_k}^H \mathbf{H}_k\mathbf{F}_{\mathrm{RF}}\mathbf{f}_{\mathrm{BB}_k}|^2 }{ \frac{P}{K}\sum_{n \neq k}|\mathbf{w}_{\mathrm{RF}_n}^H \mathbf{H}_n\mathbf{F}_{\mathrm{RF}}\mathbf{f}_{\mathrm{BB}_n} |^2 + \sigma^2}\bigg| 
	\end{align}
	and the  achievable sum-rate of the system is given by $\bar{R}=\sum_{k=1}^{K}R_k$.


	\section{Problem Formulation}
	\label{sec:ProbFormulation}
	The principal aim in this work is to design the hybrid precoder and combiners $\mathbf{F}_{\mathrm{BB}}$, $\mathbf{F}_{\mathrm{RF}}$ and $\{\mathbf{w}_{\mathrm{RF}_k}\}_{k=1}^K$ in the presence of imperfect channel data by maximizing the sum-rate. Specifically, we  first develop an algorithm to compute the hybrid precoders which maximizes the sum-rate, and then a deep network is designed such that the hybrid precoders are predicted by feeding the network with imperfect CSI. 
	
	In a nutshell, the proposed DL framework provides a nonlinear mapping from the channel matrix $\mathbf{H}$ to the analog beamformers $\mathbf{F}_{\mathrm{RF}}$ and $\{\mathbf{w}_{\mathrm{RF}_k}\}_{k=1}^K$. {\color{black}The label generation process depends on the channel model which is not required for updating the network parameters in the training stage.} Hence, CNN-MIMO can also be used for various channel models in mmWave systems \cite{channelModelSurvey}. Given that our main focus is hybrid beamforming, in this work, we use the block-fading channel model  due to the simplistic structure of channel matrix model and rate computation~\cite{deepLearningChannelAndDOAEst,deepChannelEstBeamspace1,deepChannelEstBeamspace2}. The application of DL to other channel models is the topic of ongoing research.

	The estimation process of the channel matrix of the users is a challenging task, especially in the case of a large number of antennas taking place in massive MIMO systems \cite{channelEstLargeArrays,channelEstimation1}. In addition, since the coherence time of the channel is very short in the mmWave massive MIMO scenario, the parameters related to the channel characteristics change greatly in a short time \cite{coherenceTimeRef}. To obtain a robust precoding performance, we feed the deep network with several channel realizations which are corrupted by synthetic noise in the training stage which is an offline process. Hence, in the testing stage when the network predicts the precoder weights, the network does not necessarily require the perfect CSI \cite{elbirDL_COMML}. We show, through simulations, that the proposed approach can handle the corrupted channel matrix case and exhibits satisfactory performance regarding the achievable sum-rate.
	
	{\color{black}The main stages of the proposed DL framework are label generation, training, and prediction. In the following section, we first discuss how the labels are obtained from the channel data. Then, in Section~\ref{sec:DL}, we present the details of the training and the prediction stages. }
	
	\section{Hybrid Precoding Design In Multi-User MIMO Systems}
	\label{sec:HD_Design}
	In order to design the network and training data, we first need to solve the hybrid precoding problem and obtain the labels of the training data samples. For this reason, {\color{black} we first develop an exhaustive search algorithm that visits all precoder and combiner combinations in the feasible sets $\mathcal{F}$ and $\mathcal{W}$ such that the sum-rate in (\ref{eq:Rate}) is maximized.} Then, we solve the exhaustive search problem in an offline manner to obtain the training data inputs and labels. The advantage of using a DL approach is the reduction  of the computation time of the hybrid precoding design problem and obtain near-optimum performance that can be obtained from an exhaustive search.
	
	We  start by formulating the optimization problem for hybrid precoding in the multi-user scenario as 
	\begin{align}
	\label{eq:Problem1}
	\{ \hat{\mathbf{F}}_{\mathrm{BB}}, \hat{\mathbf{F}}_{\mathrm{RF}}, \hat{\mathbf{W}}_{\mathrm{RF}} \} = &\argmax_{ {\mathbf{F}}_{\mathrm{BB}}, {\mathbf{F}}_{\mathrm{RF}}, {\mathbf{W}}_{\mathrm{RF}}} \bar{R} \nonumber \\
	\subjectto ~~~~~
	&\mathbf{F}_{\mathrm{RF}} \in \mathcal{F} , \hspace{3pt} \mathbf{W}_{\mathrm{RF}} \in \mathcal{W},\nonumber \\
	&\|\mathbf{F}_{\mathrm{RF}}\mathbf{F}_{\mathrm{BB}}\|_\mathrm{F}^2 = K,
	\end{align}
	where $\mathbf{W}_{\mathrm{RF}}=[\mathbf{w}_{\mathrm{RF}_1},\mathbf{w}_{\mathrm{RF}_2},\dots,\mathbf{w}_{\mathrm{RF}_K}]$ denotes the analog combiner of all users while $\mathcal{F}$  and $\mathcal{W}$ are the feasible sets of the precoder and combiners. In practice, both $\mathcal{F}$ and $\mathcal{W}$ are composed of the steering vectors $\mathbf{a}_\mathrm{T}(\Theta_\mathrm{T}^{(l,k)})$ and $\mathbf{a}_\mathrm{R}(\Theta_\mathrm{R}^{(l,k)})$, $\forall l,k$ with quantized phases, respectively. Specifically, the array response sets are selected as
	\begin{align}
	\mathcal{F} = \{Q(\mathbf{a}_\mathrm{T}(\Theta_\mathrm{T}^{(1,1)})),\dots,Q(\mathbf{a}_\mathrm{T}(\Theta_\mathrm{T}^{(L,K)})) \},
	\end{align} 
	and 
	\begin{align}
	\mathcal{W} = \{Q(\mathbf{a}_\mathrm{R}(\Theta_\mathrm{R}^{(1,1)})),\dots,Q(\mathbf{a}_\mathrm{R}(\Theta_\mathrm{R}^{(L,K)})) \},
	\end{align}
	where $Q(\cdot)$ denotes the phase quantization operator as mentioned before.
	
	In the exhaustive search algorithm, it is desired to visit all possible combinations of the elements in the feasible sets $\mathcal{F}$  and $\mathcal{W}$ to achieve near-optimum performance. For this reason, we design new feasible sets $\mathbb{F}$ and $\mathbb{W}$ which include all precoder and combiner combinations. The search algorithm visits all the nodes in  the direction set
	\begin{align}
	\mathbb{D} = [0,\frac{2\pi}{\bar{L}},\frac{4\pi}{\bar{L}},\dots, \frac{(\bar{L}-1)2\pi}{\bar{L}}],
	\end{align}
	where $|\mathbb{D}| = \bar{L}$. By  assuming that the BS receives $\bar{L}$ paths from each user,  the $k$th column of $\mathbf{F}_{\mathrm{RF}}$ can take $\bar{L}$ different values, i.e., $\{Q(\mathbf{a}_\mathrm{T}(\Theta_\mathrm{T}^{(l,k)}))\}_{l=1}^{\bar{L}}$. If we generalize it for all users, we have $Q_{F}=\bar{L}^K$ possible candidates to design $\mathbf{F}_{\mathrm{RF}}$. Thus, we  define a new  set as 
	\begin{align}
	\mathbb{F} = \{ \mathbb{F}_1,\mathbb{F}_2,\dots,\mathbb{F}_{Q_F} \},
	\end{align}
	where $\mathbb{F}_{q_F}\in \mathbb{C}^{N_\mathrm{T}\times K}$ is given by
	\begin{align}
	\mathbb{F}_{q_F}\! =\! [Q(\mathbf{a}_\mathrm{T}&(\Theta_\mathrm{T}^{(l_1,1)})\!), Q(\mathbf{a}_\mathrm{T}(\Theta_\mathrm{T}^{(l_2,2)})\!),\dots,Q(\mathbf{a}_\mathrm{T}(\Theta_\mathrm{T}^{(l_K,K)})\!)]\nonumber
	\end{align}
	with the indices for each user given by $l_1,l_2,\dots,l_K=1,\dots,\bar{L}$. Hence, we have $q_F = 1,\dots, \bar{L}^K$ which denotes the precoder candidates for $K$ users. In a similar way, the set for the analog combiners is defined as $\mathbb{W} = \{\mathbb{W}_1,\mathbb{W}_2,\dots,\mathbb{W}_{Q_W}  \}$ where $\mathbb{W}_{q_W} \in \mathbb{C}^{N_\mathrm{R}\times K}$ is given by
	\begin{align}
	\mathbb{W}_{q_W} \!=\! [Q(\mathbf{a}_\mathrm{R}(\Theta_\mathrm{R}^{(l_1,1)})\!),Q(\mathbf{a}_\mathrm{R}(\Theta_\mathrm{R}^{(l_2,2)})\!),\dots,Q(\mathbf{a}_\mathrm{R}(\Theta_\mathrm{R}^{(l_K,K)})\!)]\nonumber
	\end{align}
	with $\mathbf{w}_{\mathrm{RF}_k}$  selected from the $k$th column of $\mathbb{W}$, i.e., $Q(\mathbf{a}_\mathrm{R}(\Theta_\mathrm{R}^{(l_k,k)}))$.
	
	Once the analog precoders are selected from the sets $\mathbb{F}$ and $\mathbb{W}$, the effective channel $\mathbf{H}_{q_F,q_W}^{\text{eff}}\in \mathbb{C}^{K\times N_\mathrm{T}^{RF}}$ is given by 
	\begin{align}
	\mathbf{H}_{q_F,q_W}^{\text{eff}} = \left( \begin{array}{c}
	\mathbf{h}_{q_F,q_W,1}^{\text{eff}}  \\
	\mathbf{h}_{q_F,q_W,2}^{\text{eff}}  \\
	\vdots   \\
	\mathbf{h}_{q_F,q_W,K}^{\text{eff}}
	\end{array} \right),
	\end{align}
	where the corresponding effective channel for each user can be calculated as
	\begin{align}
	\mathbf{h}_{q_F,q_W,k}^{\text{eff}} = [\mathbb{W}_{q_W}]_{:,k}^H\mathbf{H}_k\mathbb{F}_{q_F}.
	\end{align}
	
	The baseband precoder can be given by $\mathbf{F}_{\mathrm{BB},q_F,q_W}= \big( \mathbf{H}_{q_F,q_W}^{\text{eff}}\big)^{\dagger}$  and it is normalized as $\mathbf{f}_{\mathrm{BB}_k}^{(q_F,q_W)} = \mathbf{f}_{\mathrm{BB}_k}^{(q_F,q_W)}/\|\mathbb{F}_{q_F}\mathbf{f}_{\mathrm{BB}_k}^{(q_F,q_W)}\|_\mathrm{F}$~\cite{Alkhateeb2015}. Thus, the achievable sum-rate then can be written as
	\begin{align}
	\label{eq:RateDiscrete}
	\bar{{R}}_{q_F,q_W} =& \log_2\bigg|\mathbf{I}_K + \nonumber \\ \frac{P}{K\sigma^2}& \mathbf{H}_{q_F,q_W}^{\text{eff}}\mathbf{F}_{\mathrm{BB},q_F,q_W}\mathbf{F}_{\mathrm{BB},q_F,q_W}^H \mathbf{H}_{q_F,q_W}^{\text{eff}^H}  \bigg|.
	\end{align}
	Using the sets $\mathbb{F}$ and $\mathbb{W}$, the optimization problem in (\ref{eq:Problem1}) can be rewritten as
	\begin{align}
	\label{eq:Problem2}
	\{\bar{q}_F, \bar{q}_W \} &= \argmax_{{q}_F, {q}_W } \bar{R}_{{q}_F, {q}_W } \nonumber \\
	\subjectto~~~~~ 
	&\mathbf{F}_{\mathrm{RF}} = \mathbb{F}_{q_F},\mathbf{w}_{\mathrm{RF}_k} = [\mathbb{W}_{q_W}]_{:,k}, \nonumber \\
	&\mathbf{h}_k^{\text{eff}} = \mathbf{w}_{\mathrm{RF}_k}^H\mathbf{H}_k\mathbf{F}_{\mathrm{RF}},\nonumber \\
	&\mathbf{F}_{\mathrm{BB}}= \big( \mathbf{H}^{\text{eff}}\big)^{\dagger}, \nonumber \\
	&\mathbf{f}_{\mathrm{BB}_k} = \mathbf{f}_{\mathrm{BB}_k}/\|\mathbf{F}_{\mathrm{RF}}\mathbf{f}_{\mathrm{BB}_k}\|_\mathrm{F},
	\end{align}
	where $\bar{q}_F$ and $\bar{q}_W$ denote the indices providing the maximum sum-rate. We summarize the algorithmic steps of the proposed approach in Algorithm~\ref{alg:1HB}. Note that the proposed hybrid precoding optimization in (\ref{eq:Problem2}) is different than the one in \cite{Alkhateeb2015}, in which, not all possible combinations of the analog precoders are considered as it is done in this work. In Section \ref{sec:Sim}, we show that (\ref{eq:Problem2}) yields better results as compared to \cite{Alkhateeb2015}. The problem in (\ref{eq:Problem2}) requires to visit $Q_FQ_W$ nodes to estimate the hybrid precoders. To reduce the complexity and the need for the array responses, in the following section,  we propose a DL-based approach where we elaborate on the details of the training data generation and network architecture.

	\begin{algorithm}[t]
		\begin{algorithmic}[1]
			\caption{Hybrid precoding for Multi-user MIMO}
			\Statex {\textbf{Input:} $\{\mathbf{H}_k\}_{k=1}^K$, $\mathbb{F}$, $\mathbb{W}$, $\mathbb{D}$.}
			\label{alg:1HB}
			\Statex {\textbf{Output:} $\hat{\mathbf{F}}_{\mathrm{RF}}$, $\hat{\mathbf{W}}_{\mathrm{RF}}$.}
			\State \textbf{for}  $1 \leq q_F\leq Q_F$ \textbf{do}
			\State \indent $\mathbf{F}_{\mathrm{RF}} = \mathbb{F}_{q_F}$,
			\State  \indent \textbf{for}  $1 \leq q_W \leq Q_W$ \textbf{do}
			\State \indent \indent $\mathbf{w}_{\mathrm{RF}_k}= [\mathbb{W}_{q_W}]_{:,k}$,
			\State \indent \indent $\mathbf{h}_k^{\text{eff}} = \mathbf{w}_{\mathrm{RF}_k}^H\mathbf{H}_k\mathbf{F}_{\mathrm{RF}}$,
			\State \indent \indent $\mathbf{F}_{\mathrm{BB}}= \big( \mathbf{H}^{\text{eff}}\big)^{\dagger},$
			\State \indent \indent $\mathbf{f}_{\mathrm{BB}_k} = \mathbf{f}_{\mathrm{BB}_k}/\|\mathbf{F}_{\mathrm{RF}}\mathbf{f}_{\mathrm{BB}_k}\|_\mathrm{F},$
			\State \indent \indent \textbf{Compute} $\bar{R}_{q_F,q_W}$ as in (\ref{eq:RateDiscrete}).
			\State \indent \textbf{end for} $q_W$,
			\State \textbf{end for} $q_F$,
			\State $\{\bar{q}_F,\bar{q}_W\} = \arg \max_{q_F,q_W}\bar{R}_{q_F,q_W}$.
			\State $\hat{\mathbf{F}}_{\mathrm{RF}}= \mathbb{F}_{\bar{q}_F}$ and  $\hat{\mathbf{W}}_{\mathrm{RF}}= \mathbb{W}_{\bar{q}_W}$.
		\end{algorithmic}
	\end{algorithm}
	\begin{figure*}[t]
		\centering
		\subfloat{\includegraphics[draft=false,scale=0.7]{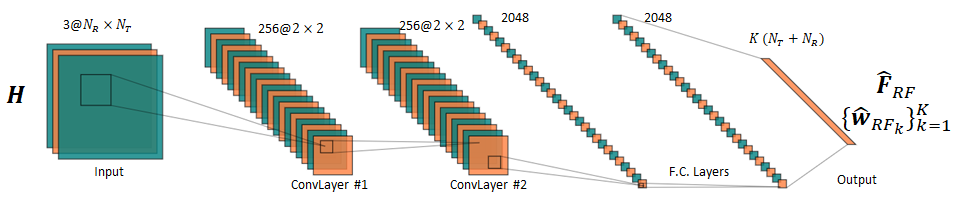} } \\
		\subfloat{\includegraphics[draft=false,scale=0.7]{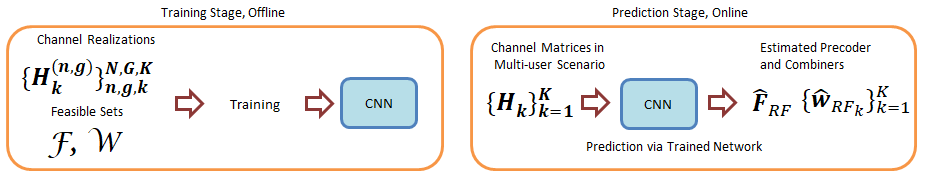} } 
		\caption{ (Top) The proposed network architecture. The input is the channel matrix of any user in the network and the output is the corresponding analog precoder and combiners. {\color{black}(Bottom) The diagram for the training and prediction stage of the proposed DL framework.}  }
		\label{fig_Network}
	\end{figure*}

	\section{Learning-Based Hybrid precoding}
	\label{sec:DL}
	In this part, we present our DL framework for hybrid precoding design. The proposed network architecture is illustrated in Fig.~\ref{fig_Network}. The CNN-MIMO architecture consists of ten layers and it accepts an input data  of size ${N_\mathrm{R}\times N_\mathrm{T}\times 3}$ while it yields a ${K(N_\mathrm{T}+ N_\mathrm{R}})\times 1$ vector at the output. 
	The overall network architecture of CNN-MIMO can be represented by the function $\mathbf{\Pi}(\cdot): \mathbb{R}^{N_\mathrm{R}\times N_\mathrm{T}\times 3} \rightarrow \mathbb{R}^{K(N_\mathrm{R} + N_\mathrm{T})}$. Let us define the arithmetic operation of the $i$th layer in the network with $f^{(i)}(\cdot)$, then the representation of the overall network can be given as
	\begin{align}
	\mathbf{\Pi}(\mathbf{X}) = f^{(10)}\big( f^{(9)} (\cdots   f^{(1)}( \mathbf{X})  \cdots)\big) = \mathbf{z},
	\end{align}
	where each layer has certain task described above and we explicitly show the arithmetic operations for fully connected layers are convolutional layers in the sequel.

	Let $\bar{\bf W} \in \mathbb{R}^{C_{x}\times C_{y}}$ be the weights of a fully connected layer in the network with input $\bar{\bf x}\in \mathbb{R}^{C_x}$ and output $\bar{\bf y}\in \mathbb{R}^{C_y}$. The $c_y$th element of the output of the layer can be given by the inner product
	\begin{align}
	\bar{\bf y}_{c_y} = \langle \bar{\bf W}_{c_y}, \bar{\bf x} \rangle = \sum_{i} {[\bar{\bf W}}]_{c_y,i}^T \bar{ \bf x}_i ,
	\end{align}
	for $c_y = 1,\dots, C_y$ and  $\bar{\bf W}_{c_y}$ is the $c_y$th column vector of $\bar{\bf W}$.

	For a convolutional layer, define $\bar{\bf X} \in \mathbb{R}^{d_{x}\times d_{x}\times C_x}$ and $\bar{\bf Y} \in \mathbb{R}^{d_{y}\times d_y\times C_{y}}$ as the feature maps and output of a convolutional layer, respectively. Let us also define  $d_x \times d_y$ as the size of the convolutional kernel, and $C_x \times C_y$ as the size of the response of convolutional layer for each feature map. 
	Then, the response of a convolutional layer becomes
	\begin{align}
	\bar{\bf Y}_{p_y,c_y} = \sum_{p_k,p_x} \langle \bar{\bf W}_{c_y,p_k}, \bar{\bf X}_{p_x} \rangle,
	\end{align}
where $\bar{\bf Y}_{p_y,c_y}$ is the response for the 2-D spatial region $p_y$ in the $c_y$th channel of the feature maps, $\bar{\bf W}_{c_y,p_k}\in \mathbb{R}^{C_x}$ denotes the weights of the $c_y$th convolutional kernel, and $\bar{\bf X}_{p_x} \in \mathbb{R}^{C_x}$ is the input feature map at spatial position $p_x$. Hence we define $p_x$ and $p_k$ as the 2-D spatial positions in the feature maps and convolutional kernels, respectively \cite{quantizedCNN_Unified}.

	\subsection{Training Data Generation}
	In order to train the network, we prepare a training dataset for several channel realizations. We generate $N$ different channel realizations for $K$ users. Next, each of these channel matrices are corrupted by a synthetic noise for $G$ realizations. The noise is added to each term in the channel matrix and we define the SNR for the training data generation as $\text{SNR}_{\text{TRAIN}} =   20\log_{10}(\frac{|[\mathbf{H}_k^{(n,g)}]_{i,j}|^2}{\sigma_{\text{TRAIN}}^2})$,
	where $\sigma_{\text{TRAIN}}^2$ is the variance of synthetic noise. Note that $[\mathbf{H}_k^{(n,g)}]_{i,j}$ denotes the $(i,j)$th entry of the $k$th channel matrix for the $(n,g)$th realization with $n=1,\dots,N$ and $g=1,\dots,G$. 
	
	The input of the network consists of three channels. In the first channel, the absolute values of the entries in the  channel matrix are used. The second and the third channels include the real and imaginary parts of the channel matrix, respectively. This approach provides good features for the solution of the problems \cite{elbirQuantizedCNN2019}. Specifically, let $\mathbf{X}\in \mathbb{R}^{N_\mathrm{R}\times N_\mathrm{T}\times 3}$ be the input of the network, then, for a channel matrix $\mathbf{H}\in \mathbb{C}^{N_\mathrm{R}\times N_\mathrm{T}}$, the first channel of the input is given by $[[\mathbf{X}]_{:,:,1}]_{i,j} = |[\mathbf{H}]_{i,j}|$. The second and the third channels are given by $[[\mathbf{X}]_{:,:,2}]_{i,j} = \operatorname{Re}\{[\mathbf{H}]_{i,j}\}$ and $[[\mathbf{X}]_{:,:,3}]_{i,j} = \operatorname{Im}\{[\mathbf{H}]_{i,j}\}$, respectively.

	The output of the network is composed of the analog precoder and combiners. Let $\mathbf{z}\in \mathbb{R}^{N_\mathrm{T}K + N_\mathrm{R} K}$ be a real valued vector, then we design the output as
	\begin{align}
	\mathbf{z} = [\angle\{\text{vec}({\mathbf{F}}_{\mathrm{RF}})^T\},\angle\{\text{vec}({\mathbf{W}}_{\mathrm{RF}})^T\}]^T,
	\end{align}
	where $\mathbf{F}_{\mathrm{RF}}\in \mathbb{C}^{N_\mathrm{T}\times K}$  and $\mathbf{W}_{\mathrm{RF}}\in \mathbb{C}^{N_\mathrm{R}\times K}$.
	Hence the input-output pair of the network is $(\mathbf{X},\mathbf{z})$. We summarize the data generation process in Algorithm~\ref{alg:algorithmTraining}. The total number of inputs is $T=NGK$ for $K$ users. Note that the input data is composed of each user channel information as in lines $7-12$ of Algorithm~\ref{alg:algorithmTraining} and we record the analog precoder and combiner associated with each user channel. Note also that the same analog precoders are used for all noisy channel realizations. This is to introduce synthetic noise in the input dataset to make the network robust against the corrupted channel data \cite{elbirQuantizedCNN2019,elbirIETRSN2019}.

	\begin{algorithm}[ht]
		\begin{algorithmic}[1]
			\caption{Training data generation for CNN-MIMO. }
			\Statex {\textbf{Input:} $N$,  $G$, $K$,  SNR$_{\text{TRAIN}}$}.
			\label{alg:algorithmTraining}
			\Statex {\textbf{Output:} Training data $\mathcal{D}_{\text{TRAIN}}$.}
			\State Generate $N$ different realizations of the  multi-user MIMO scenario with channel matrices $\{\mathbf{H}_k^{(n)}\}_{n=1}^N$ and corresponding feasible sets $\{\mathbb{F}^{(n)}\}_{n=1}^N$, $\{\mathbb{W}^{(n)}\}_{n=1}^N$ $\forall k$.
			\State Initialize with $t\hspace{-3pt}=\hspace{-3pt}1\hspace{-3pt}$ while the dataset length is $T=NGK$.
			\State   \textbf{for}  $1 \leq n \leq N$ \textbf{do}
			\State  \indent \textbf{for}  $1 \leq g \leq G$ \textbf{do}
			\State \indent  $[\mathbf{H}_k^{(n,g)}]_{i,j} \sim \mathcal{CN}([\mathbf{H}_k^{(n)}]_{i,j},\sigma_{\text{TRAIN}}^2)$.
			\State \indent  Using $\mathbf{H}_k^{(n,g)}$, $\mathbb{F}^{(n)}$, $\mathbb{W}^{(n)}$ in Algorithm~\ref{alg:1HB}, find\par \indent   $\hat{\mathbf{F}}_{\mathrm{RF}}^{(n,g)}$ and $\hat{\mathbf{W}}_{\mathrm{RF}}^{(n,g)}$ using $\bar{q}_F^{(n,g)}$ and $\bar{q}_W^{(n,g)}$.
			\State \indent \textbf{for}  $1 \leq k \leq K$ \textbf{do}
			\State   \indent \indent$[[\mathbf{X}^{(t)}]_{:,:,1}]_{i,j} = |[\mathbf{H}_k^{(n,g)}]_{i,j}|$.
			\State  \indent\indent $[[\mathbf{X}^{(t)}]_{:,:,2}]_{i,j}=\operatorname{Re} \{[\mathbf{H}_k^{(n,g)}]_{i,j}\}$ .
			\State  \indent\indent $[[\mathbf{X}^{(t)}]_{:,:,3}]_{i,j} = \operatorname{Im}\{[\mathbf{H}_k^{(n,g)}]_{i,j}\}$ $\forall ij$.
			\State  \indent\indent $\mathbf{z}^{(t)} = [\angle\{\text{vec}(\hat{\mathbf{F}}_{\mathrm{RF}}^{(n,g)})^T\},\angle\{\text{vec}(\hat{\mathbf{W}}_{\mathrm{RF}}^{(n,g)})^T\}]^T$.
			\State \indent \indent Construct the input-output pair $(\mathbf{X}^{(t)},\mathbf{z}^{(t)} )$.
			\State \indent\indent $t = t+1$.
			\State \indent\textbf{end for} $k$,
			\State \indent\textbf{end for} $g$,	
			\State \textbf{end for} $n$,
			\State Training data for CNN-MIMO is obtained from the  collection of the input-output pairs as \par \noindent $\mathcal{D}_{\text{TRAIN}} = \big((\mathbf{X}^{(1)}, \mathbf{z}^{(1)}),(\mathbf{X}^{(2)}, \mathbf{z}^{(2)}),\dots, (\mathbf{X}^{(T)}, \mathbf{z}^{(T)})\big).$
		\end{algorithmic}
	\end{algorithm}

	\subsection{Network Architecture}
	\label{sec:training}
	The proposed network shown in Fig.~\ref{fig_Network} is composed of ten layers. The first layer is the \textcolor{black}{input layer accepting the channel matrix data of size $ N_\mathrm{R}\times N_\mathrm{T}\times 3$ which denotes $3$ "channels", each of which has  size equal to $N_\mathrm{R}\times N_\mathrm{T}$.} The second and the fourth layer are the convolutional layers with 256 filters of size $2\times 2$ to extract the features hidden in the input data.  We feed the network with the real and imaginary parts of the channel data which provides a large number of features  \cite{elbirDL_COMML,elbirIETRSN2019} to be handled to help the network map and learn the input data in accordance with their label data. After each convolutional layer, there is a normalization layer to normalize the output and provide better convergence. The sixth and eighth layers are fully connected layers with 2048 units, respectively. There are dropout layers after the fully connected layers (the seventh and ninth layers) with a 50\% probability. The dropout layers make the network non-dependent on the initial weights. The output layer is the regression layer  with $K(N_\mathrm{R} + N_\mathrm{T})$ units which include the phase information of the analog precoders.  {\color{black} In order to obtain the network parameters such as the number of layers, number of filters and kernel sizes, we have conducted a hyperparameter tuning process to achieve the sufficiently good network accuracy and sum-rate performance \cite{elbirDL_COMML,elbirIETRSN2019,deepLearningScience}. } {\color{black} The current network architecture with a kernel size $2 \times 2$ is one possible solution of the considered problem with similar/same performance with network structures having different kernels. In other words, although different kernel sizes can also be used for this problem, in this work, we  have first considered a hyperparameter tuning process providing the sufficient performance for the considered scenario with less computational complexity \cite{elbirDL_COMML,elbirIETRSN2019,deepLearningScience}.}
	
	\textcolor{black}{The computational workload of a CNN is the result of intensive use of arithmetic operations in its layers. Most of the operations occur on the convolutional parts of the network. Hence, convolutional layers are responsible for more than 90\% of the execution time during the inference \cite{alexNet}. Conversely to computations, most of the CNN weights are included on the fully connected layers which require approximately 90\% of memory due to a large number of weights \cite{alexNet}. Hence, the complexity of CNN is directly proportional to the number of parameters and the number of layers. The layers of the proposed CNN structure are described above and the number of parameters can be calculated as $C^2\big(2N_{cv} (wh +1) + ([N_{fc_1}+1] + [N_{fc_2}+1])\cdot \frac{50}{100}\big)$  \cite{alexNet}. Here, $C=3$ corresponds to the number of channels, $w=h=2$ is the filter size, and $N_{cv}=256$ is the number of filters in both convolutional layers. The variables $N_{fc_1}=N_{fc_2} = 2048$ describe the number of units in the fully connected layers for $50\%$ dropout probability. Hence, the CNN-MIMO structure in Fig.~\ref{fig_Network} has $41481$ parameters.}

	\subsection{Training}
	\label{sec:Training}
	The CNN structure in Fig.~\ref{fig_Network} is realized and trained in MATLAB on a PC with a single GPU and a  768-core processor.  We have used the stochastic gradient descent algorithm with momentum 0.9 \cite{bishop2006pattern} and  updated the network parameters with learning rate $0.005$ and mini-batch size of $500$ samples for 100 epochs. As a loss function, we used the MSE given by $\mathcal{L} =  \frac{1}{T}\sum_{t=1}^{T} \big( \mathbf{z}^{(t)}- f(\mathbf{X}^{(t)})  \big)^2$
	where $f(\mathbf{X})$ is a function of the input data $\mathbf{X}$, which represents the nonlinear transformation achieved by the network \cite{deepLearningScience}. 
	
	To train the proposed CNN structure, $N=500$ different multi-user scenarios are realized with $K=3$ users (1500 channel realizations in total) as in Algorithm~\ref{alg:algorithmTraining}. For each channel matrix, AWGN is added for different powers of SNR$_{\text{TRAIN}}\in \{15,20,25\}$dB with $G=100$ to account for different channel characteristics. The use of multiple SNR$_{\text{TRAIN}}$ levels provides a wide range of corrupted data in the training which improves the learning and robustness of the network. Hence, the total size of the training data is $N_\mathrm{R}\times N_\mathrm{T}\times3 \times 450000$. In the training process, $80\%$ and $20\%$ of all generated data  are selected as the training and validation datasets, respectively. The validation aids in hyperparameter tuning during the training phase to avoid the network simply memorizing the training data rather than learning general features for accurate prediction with new data. The validation data is used to test the performance of the network in the simulations for $J_\mathrm{T}=100$ Monte Carlo trials. In order to prevent the similarity between the test data and the training data we also add synthetic noise to the test data where the SNR during testing is defined similar to SNR$_\text{TRAIN}$ as SNR$_\text{TEST} = 20\log_{10}(\frac{|[\mathbf{H}]_{i,j}|^2}{\sigma_{\text{TEST}}^2})$. The number of grid points is selected as $\bar{L}=60$ for azimuth and $\bar{L}=20$ for elevation angular sectors in Algorithm~\ref{alg:1HB}. In addition, the propagation environment is modeled with $L=10$ paths from the users and all the user directions, i.e., all the azimuth and elevation angles, are uniform randomly selected from the intervals $\phi \in [-30^{\circ},30^{\circ}]$  and $\theta \in [-20^{\circ},20^{\circ}] $, respectively \cite{mimoRHeath}. {\color{black} We  use sectorized angular range by selecting the antenna gains $g_\mathrm{R}(\Theta_\mathrm{R}^{\color{black}(l,k)}), g_\mathrm{T}(\Theta_\mathrm{T}^{\color{black}(l,k)})$  as unity for these angular ranges and zero otherwise   to provide a sectorized angular interval  increasing the beamforming gain and reducing interference and provide increased beamforming gain \cite{mimoHybridLeus1}.} Hence, the training data  includes a large number of scenarios where the users are randomly located. For each scenario, the corresponding precoder and combiners are obtained by Algorithm~\ref{alg:1HB}. 
	
{\color{black}The training stage takes about 5 hours for $T=450000$ samples. This process includes both the labeling and the input data generation. Note that the training stage is performed only once. Then, in the prediction stage, it takes only milliseconds to estimate the hybrid precoders as demonstrated in the simulations (please see Table~\ref{tableCompTimes}). Hence, the proposed approach, providing  high data rate and low latency, is quite attractive since it meets the  5G requirements.}
	
{\color{black} The trained network can work for different parameters such as the number of users\footnote[1]{\color{black} When the network is trained for $K_\mathrm{TRAIN}$ users, the output size of the network is $\mathbf{z} \in \mathbb{R}^{N_\mathrm{T}K_\mathrm{TRAIN} + N_\mathrm{R} K_\mathrm{TRAIN}}$. Then we can use the trained network for hybrid beamforming when there are $K \leq K_\mathrm{TRAIN}$ users by substituting  network output of size $N_\mathrm{T}K + N_\mathrm{R} K \times 1$ corresponding to those $K$ users.} $K$, number of paths $L$, SNR$_\mathrm{TEST}$ and SNR$_\mathrm{TRAIN}$ which motivates the practical implementation of the proposed DL framework. The proposed CNN structure requires to be retrained if there is a change in the parameters like $N_\mathrm{T}$, $N_\mathrm{R}$, $N_\mathrm{T}^\mathrm{RF}$, which directly dictate the input and output dimensions of the deep network. The performance of the network also depends on the angular interval selected in $\mathbb{D}$ when designing the feasible sets $\mathcal{F}$ and $\mathcal{W}$ as well as the antenna gains obtaining sectorized angular intervals.}

\subsection{Prediction}
{\color{black}
Once the CNN-MIMO is trained offline as demonstrated in Fig.~\ref{fig_Network}, it can be used for the prediction of the hybrid beamformers.  In order to generate the test data in the prediction stage, we have picked users randomly from the validation data and the synthetic noise is also added to the test data with SNR$_\text{TEST}$ to eliminate the similarity between the test and training datasets. The corrupted channel data of each user is fed to the network and the analog precoders are predicted from the output layer of the network. Then, their phases are quantized in $[0, 2\pi]$ with $2^B$ discrete points. Specifically, the values of the quantized phases belong in the set $\{ \frac{2\pi b}{2^B}\}_{b=1}^{2^B}$	to allow the realization of the analog precoder and combiners in a hardware-efficient manner.}

	\section{Numerical Simulations}
	\label{sec:Sim}
	In this section, we present the performance of the proposed method, CNN-MIMO, via several experiments {\color{black} where we train the network with the parameters described in Section \ref{sec:training} such as $N=500$, $K=3$, $G=100$, SNR$_\text{TRAIN}=\{15,20,25\}$ dB, learning rate $0.005$, batch size $500$ and number of epochs $100$.} We compare the performance of CNN-MIMO with state-of-the-art hybrid precoding techniques such as the manifold optimization (MO) \cite{hybridBFAltMin}, the low-resolution hybrid beamforming (LRHB) \cite{hybridBFLowRes}, SOMP \cite{mimoRHeath} and the two-stage hybrid beamforming (TS-HB) algorithm \cite{Alkhateeb2015}.  While manifold optimization and SOMP were proposed for a single-user scenario, we adapt the algorithms for the multi-user case by using the same strategy for interference cancellation as in \cite{Alkhateeb2015}. CNN-MIMO is also compared with the DL-based approach MLP proposed in \cite{mimoDeepPrecoderDesign}. MLP is designed as described in \cite{mimoDeepPrecoderDesign} but adapted for the multi-user scenario with the same training data used for CNN-MIMO. As another benchmark and  denoted as "No interference" in the simulations, we present the performance of fully-digital beamforming and combining where the interference is completely eliminated. In addition, the performance plot of the precoders used in the test data (obtained from Algorithm~\ref{alg:1HB}) is indicated as "Algorithm 1" in the experiments.

	\begin{figure}[ht]
		\centering
		{\includegraphics[draft=false,width=.35\textheight,height=.25\textheight]{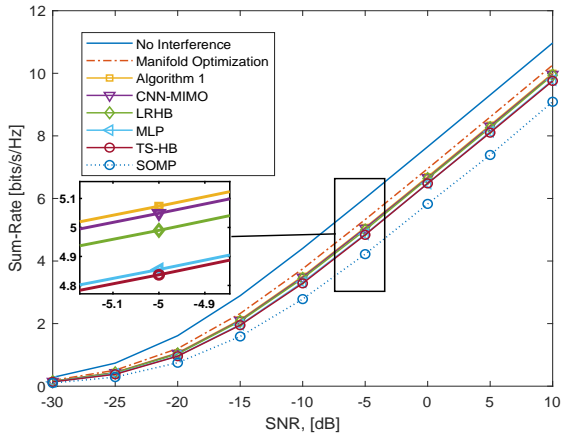} } 
		\caption{Sum-rate versus SNR ($N_\mathrm{T}=36$, $N_\mathrm{R}=9$, $K=3$, $B=3$ and  SNR$_\text{TEST}=20$ dB).  }
		\label{fig_SNR_TEST}
	\end{figure}
	\begin{figure}[ht] 
		\centering
		\subfloat[]{\includegraphics[draft=false,width=.35\textheight,height= .24\textheight]{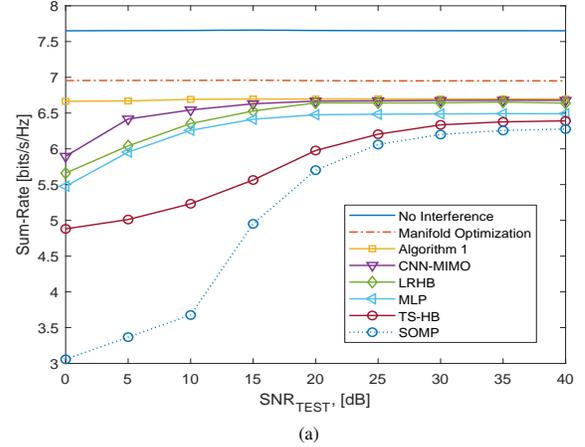} } \\
		\subfloat[]{\includegraphics[draft=false,width=.35\textheight,height= .24\textheight]{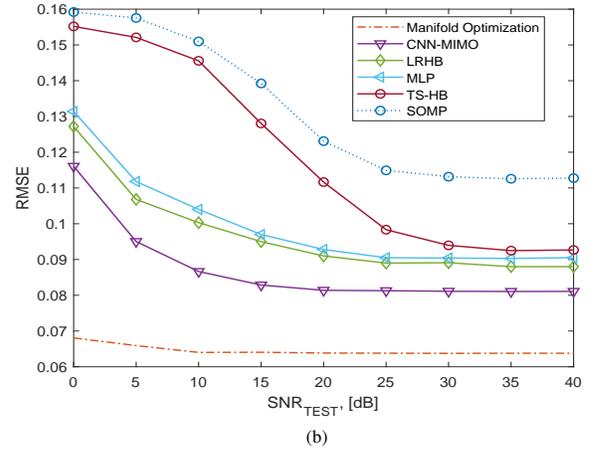} } \\
		\subfloat[]{\includegraphics[draft=false,width=.35\textheight,height= .24\textheight]{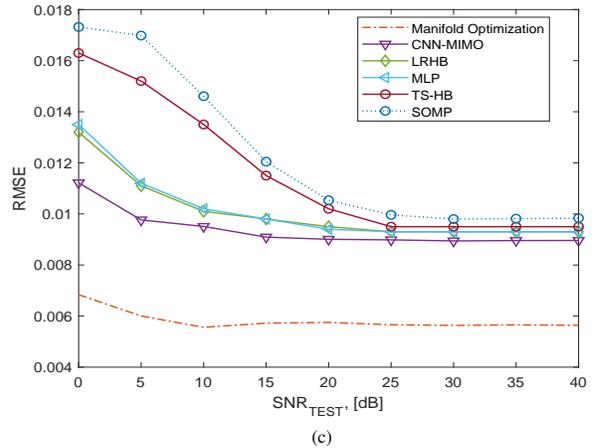} } 
		\caption{Performance comparison for corrupted channel data. In (a), sum-rate versus SNR$_\text{TEST}$  is given whereas the RMSE for precoder $\mathbf{F}_{\mathrm{RF}}$ and combiner $\mathbf{W}_{\mathrm{RF}}$ are shown in (b) and (c), respectively ($N_\mathrm{T}=36$, $N_\mathrm{R}=9$, $K=3$, $B=3$ and  SNR$=0$ dB). }
		\label{fig_SNR_DATA_TEST}
	\end{figure}
	
	In Fig.~\ref{fig_SNR_TEST}, we present the achievable sum-rate performance of the algorithms with respect to different SNR levels. 	{\color{black} The design parameters of CNN-MIMO are given in Section IV-B. Moreover, we select the number of antennas per BS and per user as $N_\mathrm{T}=36$, $N_\mathrm{R}=9$, respectively.} Synthetic noise is added to both the channel matrices and the array responses with SNR$_\text{TEST}=20$ dB and $B=3$ quantization bits are used. The number of users is $K=3$ and there are $L=10$ paths for each user. As a benchmark, we use the fully digital beamforming and the MO algorithm which has the best performance since it obtains near-optimum analog and baseband precoders. {\color{black} Our CNN approach follows the performance of the MO algorithm. In fact, CNN-MIMO provides the highest sum-rate as compared to the other algorithms. Notably, although LRHB is the state-of-the-art technique based on phase extraction and it is regarded as the technique having the best performance in the literature \cite{hybridBFLowRes}, we observe the outperformance of CNN-MIMO. MLP  has poorer performance due to the lack of feature extraction that is achieved by the convolutional layers in CNN-MIMO. {\color{black}In particular, the effectiveness of CNN-MIMO can be attributed to the maximization of the sum-rate by visiting all possible combinations for the analog parts at  both the receiver and transmitter side through an exhaustive search and well-trained deep network. We can point out that the ultimate performance from CNN-MIMO can be obtained if CNN-MIMO yields the output exactly the same as the labels obtained in Algorithm 1. Hence, we can say that the performance of CNN-MIMO is limited by the performance of Algorithm 1. We observe that the performance of CNN-MIMO is close to Algorithm 1 where the gap between these two is due to the corruption in the input data.} SOMP and TS-HB have poorer performance as compared to CNN-MIMO. Especially, while SOMP was initially proposed for the single-user case, we have  adapted the algorithm for the multi-user scenario where the analog precoders are designed based on the similarity between the optimum precoder and the analog precoders. As a result, SOMP does not always find the optimum weights maximizing the sum-rate \cite{elbirQuantizedCNN2019}. TS-HB algorithm has better performance than SOMP since it is based on the maximization of the sum-rate and its performance converges to the same one as SOMP when there is a single path from each user. }

	\begin{figure}[ht]
		\centering
		{\includegraphics[draft=false,width=.35\textheight,height=.24\textheight]{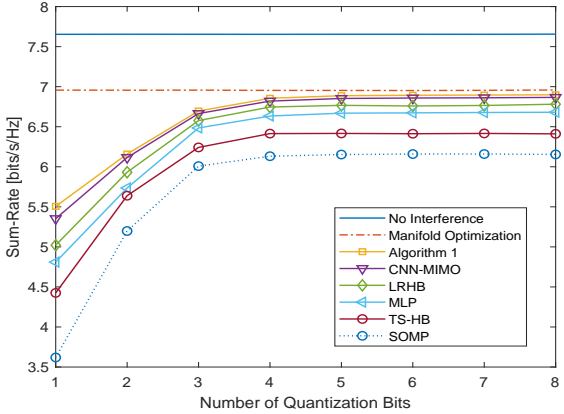} } 
		\caption{Sum-rate versus angular resolution of the analog precoders  ($N_\mathrm{T}=36$, $N_\mathrm{R}=9$, SNR$=0$ dB, SNR$_\text{TEST}=20$ dB). }
		\label{fig_Quantization_TEST}
	\end{figure}
	
	The feedback data, namely, the channel matrix $\{\mathbf{H}_k\}_{k=1}^{K}$ and the feasible array response sets $\mathcal{F}$ and $\mathcal{W}$ may not always be perfectly available. In order to evaluate the performance of the algorithms on the robustness against the corrupted feedback, we simulate the performance of the algorithms for different SNR$_\text{TEST}$ levels for the same setting as in the previous simulation. In this case, complex AWGN was added to both channel and  array response data to resemble the deviations in the feedback data. The results are presented in Fig.~\ref{fig_SNR_DATA_TEST} where we present the achievable sum-rate in Fig.\ref{fig_SNR_DATA_TEST}(a) while the RMS error on precoder $\mathbf{F}_{\mathrm{RF}}$ and combiner $\mathbf{W}_{\mathrm{RF}}$ are shown in Figs.~\ref{fig_SNR_DATA_TEST}(b) and~\ref{fig_SNR_DATA_TEST}(c), respectively. Note that Algorithm~\ref{alg:1HB} is fed with perfect CSI to demonstrate the best achievable performance. As can be seen from Fig.~\ref{fig_SNR_DATA_TEST}, CNN-MIMO is more robust against the corruption in the channel data as compared to the other methods. Note that the manifold optimization, LRHB, MLP, and CNN-MIMO are only affected by the corruption in the channel data since they automatically estimate the analog precoders, unlike SOMP and TS-HB which require the feasible sets $\mathcal{F}$ and $\mathcal{W}$ as input. As a result, the performance of TS-HB and SOMP heavily rely on the accuracy of both the channel matrix and the array response sets.  Moreover, the knowledge of channel data {\color{black} and the feasible sets $\mathcal{F}$ and $\mathcal{W}$} is only needed in the training stage of the network to obtain the labels and it is not used in the prediction stage. However, the other algorithms like SOMP and TS-HB,  require this information to solve the hybrid precoding problem. Overall, these results show the robustness of the proposed CNN-MIMO.
	
	The analog precoders are designed with discrete phase shifters with constant modulus to steer the beam in spatial precoding. To assess the performance for the phase resolution in the phase shifters, we present the sum-rate of the algorithms for different quantization resolutions where the phases of the analog precoder and combiners are quantized for $B=\{1,\dots,8\}$ bits. The results are depicted in Fig.~\ref{fig_Quantization_TEST} where we  observe that the other algorithms converge after 4 bits while, remarkably,  the proposed CNN approach achieves higher sum-rate starting from one-bit quantization.

\begin{figure}[ht]
		\centering
		\subfloat[]{\includegraphics[draft=false,width=.35\textheight,height=.24\textheight]{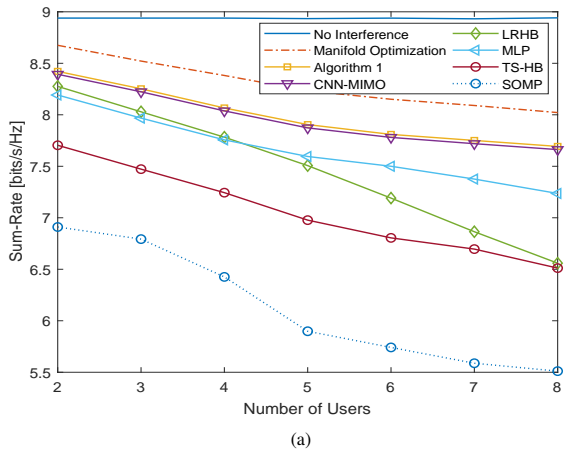} }  \\
		\subfloat[]{\includegraphics[draft=false,width=.35\textheight,height=.24\textheight]{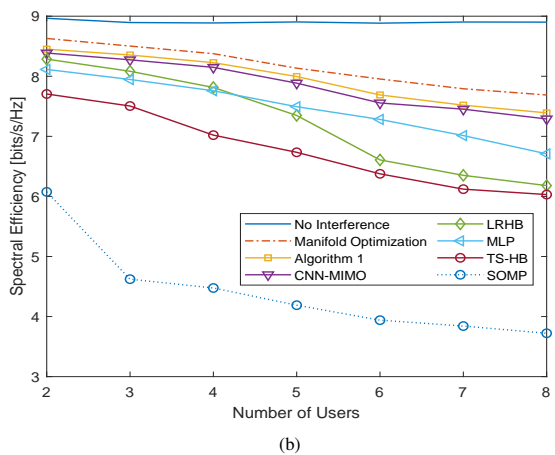} } 
		\caption{Sum-rate versus number of users. The number of paths is fixed as $L=10$ in (a), and $L$ is selected uniform randomly in the interval $[1,10]$ in (b) respectively.   ($N_\mathrm{T}=100$, $N_\mathrm{R}=9$, SNR$=0$ dB and SNR$_\text{TEST}=20$ dB). }
		\label{fig_Num_Users_TEST}
	\end{figure}

	In Fig.~\ref{fig_Num_Users_TEST}(a), the performance is evaluated for varying number of users, namely, $K\in\{2,\dots,8\}$ {\color{black}where $L=10$ is fixed}. Notably, CNN-MIMO performs better than the other algorithms. In particular,  the gap between "No interference"  and CNN-MIMO becomes larger as $K$ increases. We observe that the performance of MLP becomes better than LRHB after $K\geq 5$ and exhibits robust performance like CNN-MIMO with a certain performance loss. The main reason is that the use of training data prepared with Algorithm 1 which provides more accurate beamformers than the other algorithms. We also see that CNN-MIMO closely follows the performance of Algorithm 1. However, this gap appears due to the insufficient performance of interference cancellation. Hence, it is suggested to develop more effective algorithms to handle the interference among the users.
	
	{\color{black} In Fig.~\ref{fig_Num_Users_TEST}(b), we evaluate the performance of CNN-MIMO when the number of paths for each user is not fixed. Hence, we train the network with the same parameters except selecting $L$ uniform randomly from the interval $[1,10]$. Using varying $L$ values for different users reduces the similarity between the channel data of users and we obtain satisfactory performance of CNN-MIMO similar to the observations made when $L$ is fixed. }
	
	\begin{figure}[h]
		\centering
		{\includegraphics[draft=false,width=.35\textheight,height=.24\textheight]{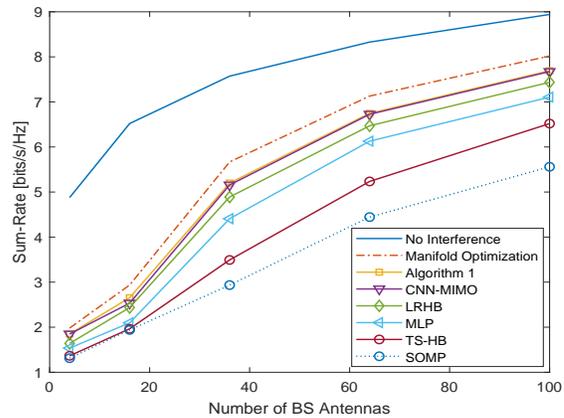} } 
		\caption{Sum-rate versus number of BS antennas ($K=3$, $N_\mathrm{R}=9$, SNR$=0$ dB and SNR$_\text{TEST}=20$ dB). }
		\label{fig_Num_BSAntennas_TEST}
	\end{figure}

	\begin{table}[h]
		\centering
		\caption{Computation Times (In seconds).}
		\label{tableCompTimes}
		\begin{tabular}{ c | c |c | c | c | c  |c}
			\hline
			\hline
			$N_\mathrm{T}$	&\hspace{-5pt}Algorithm 1 \hspace{-8pt} & \hspace{-7pt} CNN-MIMO\hspace{-4pt} & MLP& LRHB &TS-HB & SOMP\\
			\hline 
			4	&0.1061 & 0.0039 & 0.0034  & 0.0059  & 0.0093 & 0.0122\\
			\hline 
			16	&0.1164  & 0.0043 & 0.0038  & 0.0113   &  0.0103 & 0.0139\\
			\hline 
			64	&0.1175 & 0.0049 & 0.0045 & 0.0159   & 0.0108 &   0.0216\\
			\hline 
			100	&0.1242 & 0.0052 &0.0049  &  0.0318 & 0.0125  & 0.0282\\
			\hline 
			\hline
		\end{tabular}
	\end{table}   
	

	In Fig.~\ref{fig_Num_BSAntennas_TEST}, we illustrate the performance for varying number of BS antennas. As can be seen, similar observations can be obtained. Specifically,  CNN-MIMO performs better than the other algorithms. Furthermore, we  present the computation times of the algorithms for a different number of BS antennas in Table~\ref{tableCompTimes} in seconds. While the complexity of Algorithm~\ref{alg:1HB} is the highest due to the exhaustive search, DL-based approaches, i.e., CNN-MIMO and MLP have the least computation time as compared to LRHB and the rest. MLP appears slightly lower complexity than CNN-MIMO due to its less complex structure, however, it has poorer performance as was shown in the previous experiments. In addition,  regarding the complexity of TS-HB and SOMP, given its dependence on the number of elements in the feasible sets $\mathcal{F}$ and $\mathcal{W}$, it is  observed that TS-HB has less computation time than SOMP since it does not follow an OMP stage to obtain the precoders but it selects the ones with the highest channel gain from the codebook \cite{Alkhateeb2015}. {\color{black}It is also worthwhile to mention the trade-off between the computation time and the performance of CNN-MIMO. While the MO algorithm has slightly better performance than CNN-MIMO, the proposed DL framework provides a significantly faster computation of the hybrid beamformers than the MO algorithm. The complexity of MO also increases at a higher rate than that of CNN-MIMO. This observation demonstrates that CNN-MIMO is more useful in terms of computational complexity even for a very large number of antennas which is the case in 5G systems. Hence, we believe the proposed approach can be a promising technique to be used in mmWave systems where low complexity and robust performance  are required.} The run times of CNN-MIMO can be further accelerated by implementing the network in general-purpose hardware such as FPGA. For example, domain-specific architectures have been implemented in   \cite{acceleratedCNN_FPGA_300mW} for AlexNet \cite{alexNet} and VGG-16 for  real-time image classification with 194 GOP/s (billions of fixed-point OPerations per second) and consuming only 300 mW. These promising results  encourage us to develop more energy-efficient DL approaches for the problems in communications systems.
	
	%
	%
	%

	\section{Conclusions}\label{Conclusion}
	We proposed a DL framework for hybrid precoding design in multi-user mmWave MIMO systems. The proposed network architecture is a CNN which accepts as input the channel matrix of users and gives at the output the analog precoder and combiners. The proposed technique was compared with both optimization- and greedy-based approaches as well as DL-based techniques such as MLP. The effectiveness of the proposed CNN approach was evaluated through several experiments and it is shown that CNN-MIMO achieves a better performance than the state-of-the-art hybrid precoding approaches as well as less computation time. The effectiveness of CNN-MIMO can be attributed to the use of exhaustive search to obtain the best analog precoders and combiners in the training stage. In order to train the network, a large training data, with a length of nearly half a million, was used. Notably, large training data provides robust performance against the deviations in the channel data. Moreover, we showed that CNN-MIMO achieves more robust results in the presence of imperfections regarding the channel matrix and array responses.

	\balance 
	\bibliographystyle{ieeetr}
	\bibliography{IEEEabrv,references_043,Paperpile_Mar_01_BibTeX_Export}

\begin{thebibliography}{10}

\bibitem{mimoOverview}
R.~W. Heath, N.~Gonz\'{a}lez-Prelcic, S.~Rangan, W.~Roh, and A.~M. Sayeed,
  ``{An Overview of Signal Processing Techniques for Millimeter Wave MIMO
  Systems},'' {\em {IEEE} J. Sel. Topics Signal Process.}, vol.~10,
  pp.~436--453, April 2016.

\bibitem{5GwhatWillItBe}
J.~G. Andrews, S.~Buzzi, W.~Choi, S.~V. Hanly, A.~Lozano, A.~C.~K. Soong, and
  J.~C. Zhang, ``{What Will 5G Be?},'' {\em {IEEE} J. Sel. Areas Commun.},
  vol.~32, pp.~1065--1082, June 2014.

\bibitem{mimoScalingUp}
F.~Rusek, D.~Persson, B.~K. Lau, E.~G. Larsson, T.~L. Marzetta, O.~Edfors, and
  F.~Tufvesson, ``{Scaling Up MIMO: Opportunities and Challenges with Very
  Large Arrays},'' {\em {IEEE} Signal Process. Mag.}, vol.~30, pp.~40--60, Jan
  2013.

\bibitem{mmWaveKeyElements}
L.~Wei, R.~Q. Hu, Y.~Qian, and G.~Wu, ``{Key elements to enable millimeter wave
  communications for 5G wireless systems},'' {\em IEEE Wireless
  Communications}, vol.~21, pp.~136--143, December 2014.

\bibitem{mimoHybridLeus1}
A.~Alkhateeb, O.~E. Ayach, G.~Leus, and R.~W. Heath, ``{Hybrid precoding for
  millimeter wave cellular systems with partial channel knowledge},'' in {\em
  2013 Information Theory and Applications Workshop (ITA)}, pp.~1--5, Feb 2013.

\bibitem{mimoRHeath}
O.~E. Ayach, S.~Rajagopal, S.~Abu-Surra, Z.~Pi, and R.~W. Heath, ``{Spatially
  Sparse Precoding in Millimeter Wave MIMO Systems},'' {\em {IEEE} Trans.
  Wireless Commun.}, vol.~13, pp.~1499--1513, March 2014.

\bibitem{Alkhateeb2015}
A.~Alkhateeb, G.~Leus, and R.~W. Heath, ``Limited feedback hybrid precoding for
  {Multi-User} millimeter wave systems,'' {\em IEEE Trans. Wireless Commun.},
  vol.~14, pp.~6481--6494, Nov. 2015.

\bibitem{hybridBFLowRes}
Z.~{Wang}, M.~{Li}, Q.~{Liu}, and A.~L. {Swindlehurst}, ``{Hybrid Precoder and
  Combiner Design With Low-Resolution Phase Shifters in mmWave MIMO Systems},''
  {\em {IEEE} J. Sel. Topics Signal Process.}, vol.~12, pp.~256--269, May 2018.

\bibitem{mimoMU_simultaneousChannelEst}
M.~{Kokshoorn}, H.~{Chen}, Y.~{Li}, and B.~{Vucetic}, ``{Beam-On-Graph:
  Simultaneous Channel Estimation for mmWave MIMO Systems With Multiple
  Users},'' {\em {IEEE} Trans. Commun.}, vol.~66, pp.~2931--2946, July 2018.

\bibitem{mimoMU_jointAS_HB}
X.~{Zhai}, Y.~{Cai}, Q.~{Shi}, M.~{Zhao}, G.~Y. {Li}, and B.~{Champagne},
  ``{Joint Transceiver Design With Antenna Selection for Large-Scale MU-MIMO
  mmWave Systems},'' {\em {IEEE} J. Sel. Areas Commun.}, vol.~35,
  pp.~2085--2096, Sep. 2017.

\bibitem{deepLearningScience}
Y.~Lecun, Y.~Bengio, and G.~Hinton, ``Deep learning,'' {\em Nature}, vol.~521,
  no.~7553, pp.~436--444, 2015.

\bibitem{deepLearning4SignalProcessing}
D.~Yu and L.~Deng, ``Deep learning and its applications to signal and
  information processing [exploratory dsp],'' {\em {IEEE} Signal Process.
  Mag.}, vol.~28, pp.~145--154, Jan 2011.

\bibitem{elbirIETRSN2019}
A.~M. Elbir, K.~V. Mishra, and Y.~C. Eldar, ``{Cognitive radar antenna
  selection via deep learning},'' {\em IET Radar, Sonar \& Navigation},
  vol.~13, pp.~871--880(9), June 2019.

\bibitem{deepLearningWC2}
Z.~{Jiang}, S.~{Chen}, A.~F. {Molisch}, R.~{Vannithamby}, S.~{Zhou}, and
  Z.~{Niu}, ``{Exploiting Wireless Channel State Information Structures Beyond
  Linear Correlations: A Deep Learning Approach},'' {\em {IEEE} Commun. Mag.},
  vol.~57, pp.~28--34, March 2019.

\bibitem{deepLearningWC3}
M.~{Feng} and S.~{Mao}, ``{Dealing with Limited Backhaul Capacity in
  Millimeter-Wave Systems: A Deep Reinforcement Learning Approach},'' {\em
  {IEEE} Commun. Mag.}, vol.~57, pp.~50--55, March 2019.

\bibitem{mimoDLChannelEstimation}
H.~Ye, G.~Y. Li, and B.~Juang, ``{Power of Deep Learning for Channel Estimation
  and Signal Detection in OFDM Systems},'' {\em IEEE Wireless Communications
  Letters}, vol.~7, pp.~114--117, Feb 2018.

\bibitem{mimoDLChannelEstDOAEst}
H.~Huang, J.~Yang, H.~Huang, Y.~Song, and G.~Gui, ``{Deep Learning for
  Super-Resolution Channel Estimation and DOA Estimation Based Massive MIMO
  System},'' {\em {IEEE} Trans. Veh. Technol.}, vol.~67, pp.~8549--8560, Sep.
  2018.

\bibitem{mimoDLHybrid}
Y.~Long, Z.~Chen, J.~Fang, and C.~Tellambura, ``{Data-Driven-Based Analog Beam
  Selection for Hybrid Beamforming Under mm-Wave Channels},'' {\em {IEEE} J.
  Sel. Topics Signal Process.}, vol.~12, pp.~340--352, May 2018.

\bibitem{mimoDLDetection}
N.~Samuel, T.~Diskin, and A.~Wiesel, ``{Deep MIMO detection},'' in {\em {2017
  IEEE 18th International Workshop on Signal Processing Advances in Wireless
  Communications (SPAWC)}}, pp.~1--5, July 2017.

\bibitem{mimoDL_reinforcementL}
S.~{Wang}, H.~{Liu}, P.~H. {Gomes}, and B.~{Krishnamachari}, ``{Deep
  Reinforcement Learning for Dynamic Multichannel Access in Wireless
  Networks},'' {\em IEEE Transactions on Cognitive Communications and
  Networking}, vol.~4, pp.~257--265, June 2018.

\bibitem{mimoDLOverAir}
S.~D\"{o}rner, S.~Cammerer, J.~Hoydis, and S.~t.~Brink, ``{Deep Learning Based
  Communication Over the Air},'' {\em {IEEE} J. Sel. Topics Signal Process.},
  vol.~12, pp.~132--143, Feb 2018.

\bibitem{mimoDLOverAir2}
V.~Raj and S.~Kalyani, ``{Backpropagating Through the Air: Deep Learning at
  Physical Layer Without Channel Models},'' {\em {IEEE} Commun. Lett.},
  vol.~22, pp.~2278--2281, Nov 2018.

\bibitem{mimoDLCSIFeedBack}
C.~Wen, W.~Shih, and S.~Jin, ``{Deep Learning for Massive MIMO CSI Feedback},''
  {\em IEEE Wireless Communications Letters}, vol.~7, pp.~748--751, Oct 2018.

\bibitem{Raj2018-at}
V.~Raj and S.~Kalyani, ``Backpropagating through the air: Deep learning at
  physical layer without channel models,'' {\em IEEE Commun. Lett.}, vol.~22,
  pp.~2278--2281, Nov. 2018.

\bibitem{deepChannelEstBeamspace1}
P.~Dong, H.~Zhang, G.~Y. Li, N.~Naderializadeh, and I.~Gaspar, ``Deep cnn based
  channel estimation for mmwave massive mimo systems,'' {\em ArXiv},
  vol.~abs/1904.06761, 2019.

\bibitem{deepChannelEstBeamspace2}
H.~{He}, C.~{Wen}, S.~{Jin}, and G.~Y. {Li}, ``{Deep Learning-Based Channel
  Estimation for Beamspace mmWave Massive MIMO Systems},'' {\em IEEE Wireless
  Communications Letters}, vol.~7, pp.~852--855, Oct 2018.

\bibitem{mimoDeepPrecoderDesign}
H.~{Huang}, Y.~{Song}, J.~{Yang}, G.~{Gui}, and F.~{Adachi},
  ``{Deep-Learning-based Millimeter-Wave Massive MIMO for Hybrid Precoding},''
  {\em {IEEE} Trans. Veh. Technol.}, pp.~1--1, 2019.

\bibitem{deepMISO}
T.~{Lin} and Y.~{Zhu}, ``{Beamforming Design for Large-Scale Antenna Arrays
  Using Deep Learning},'' {\em arXiv e-prints}, p.~arXiv:1904.03657, Apr 2019.

\bibitem{mimoDLChannelModelBeamformingFacebook}
A.~Alkhateeb, S.~P. Alex, P.~Varkey, Y.~Li, Q.~Z. Qu, and D.~Tujkovic, ``{Deep
  Learning Coordinated Beamforming for Highly-Mobile Millimeter Wave
  Systems},'' {\em IEEE Access}, vol.~6, pp.~37328--37348, 2018.

\bibitem{elbirDL_COMML}
A.~M. {Elbir}, ``{CNN}-based precoder and combiner design in {mmWave MIMO}
  systems,'' {\em {IEEE} Commun. Lett.}, vol.~23, no.~7, pp.~1240--1243, 2019.

\bibitem{elbirQuantizedCNN2019}
A.~M. {Elbir} and K.~V. {Mishra}, ``{Joint Antenna Selection and Hybrid
  Beamformer Design using Unquantized and Quantized Deep Learning Networks},''
  {\em arXiv e-prints}, p.~arXiv:1905.03107, May 2019.

\bibitem{Yu2016-gn}
X.~Yu, J.~Shen, J.~Zhang, and K.~B. Letaief, ``Alternating minimization
  algorithms for hybrid precoding in millimeter wave {MIMO} systems,'' {\em
  IEEE J. Sel. Top. Signal Process.}, vol.~10, pp.~485--500, Apr. 2016.

\bibitem{mmWaveModel1}
E.~Torkildson, C.~Sheldon, U.~Madhow, and M.~Rodwell, ``{Millimeter-Wave
  Spatial Multiplexing in an Indoor Environment},'' in {\em 2009 IEEE Globecom
  Workshops}, pp.~1--6, Nov 2009.

\bibitem{mimoChannelModel1}
R.~M\'{e}ndez-Rial, C.~Rusu, A.~Alkhateeb, N.~González-Prelcic, and R.~W.
  Heath, ``{Channel estimation and hybrid combining for mmWave: Phase shifters
  or switches?},'' in {\em 2015 Information Theory and Applications Workshop
  (ITA)}, pp.~90--97, Feb 2015.

\bibitem{Raghavan06multi-antennacapacity}
V.~Raghavan and A.~M. Sayeed, ``{Multi-antenna capacity of sparse multipath
  channels},'' {\em IEEE TRANS. INFORM. THEORY}, 2006.

\bibitem{RappaportChannelModel}
T.~S. {Rappaport}, F.~{Gutierrez}, E.~{Ben-Dor}, J.~N. {Murdock}, Y.~{Qiao},
  and J.~I. {Tamir}, ``{Broadband Millimeter-Wave Propagation Measurements and
  Models Using Adaptive-Beam Antennas for Outdoor Urban Cellular
  Communications},'' {\em {IEEE} Trans. Antennas Propag.}, vol.~61,
  pp.~1850--1859, April 2013.

\bibitem{channelModelSurvey}
I.~A. {Hemadeh}, K.~{Satyanarayana}, M.~{El-Hajjar}, and L.~{Hanzo},
  ``{Millimeter-Wave Communications: Physical Channel Models, Design
  Considerations, Antenna Constructions, and Link-Budget},'' {\em {IEEE}
  Commun. Surveys Tuts.}, vol.~20, pp.~870--913, Secondquarter 2018.

\bibitem{deepLearningChannelAndDOAEst}
H.~Huang, J.~Yang, H.~Huang, Y.~Song, and G.~Gui, ``Deep learning for
  super-resolution channel estimation and doa estimation based massive mimo
  system,'' {\em {IEEE} Trans. Veh. Technol.}, vol.~67, pp.~8549--8560, Sept
  2018.

\bibitem{channelEstLargeArrays}
Z.~{Marzi}, D.~{Ramasamy}, and U.~{Madhow}, ``{Compressive Channel Estimation
  and Tracking for Large Arrays in mm-Wave Picocells},'' {\em {IEEE} J. Sel.
  Topics Signal Process.}, vol.~10, pp.~514--527, April 2016.

\bibitem{channelEstimation1}
J.~Wang, Z.~Lan, C.~woo Pyo, T.~Baykas, C.~sean Sum, M.~A. Rahman, J.~Gao,
  R.~Funada, F.~Kojima, H.~Harada, and S.~Kato, ``{Beam codebook based
  beamforming protocol for multi-Gbps millimeter-wave WPAN systems},'' {\em
  {IEEE} J. Sel. Areas Commun.}, vol.~27, pp.~1390--1399, October 2009.

\bibitem{coherenceTimeRef}
E.~{Bj{\"o}rnson}, L.~{Van der Perre}, S.~{Buzzi}, and E.~G. {Larsson},
  ``{{Massive MIMO in Sub-6 GHz and mmWave: Physical, Practical, and Use-Case
  Differences}},'' {\em arXiv e-prints}, p.~arXiv:1803.11023, Mar 2018.

\bibitem{quantizedCNN_Unified}
J.~{Cheng}, J.~{Wu}, C.~{Leng}, Y.~{Wang}, and Q.~{Hu}, ``Quantized {CNN}: {A}
  unified approach to accelerate and compress convolutional networks,'' {\em
  IEEE Transactions on Neural Networks and Learning Systems}, vol.~29, no.~10,
  pp.~4730--4743, 2018.

\bibitem{alexNet}
A.~Krizhevsky, I.~Sutskever, and G.~E. Hinton, ``Imagenet classification with
  deep convolutional neural networks,'' in {\em Advances in Neural Information
  Processing Systems}, pp.~1097--1105, 2012.

\bibitem{bishop2006pattern}
C.~M. Bishop, {\em Pattern Recognition and Machine Learning}.
\newblock Springer, New York, 2006.

\bibitem{hybridBFAltMin}
X.~{Yu}, J.~{Shen}, J.~{Zhang}, and K.~B. {Letaief}, ``{Alternating
  Minimization Algorithms for Hybrid Precoding in Millimeter Wave MIMO
  Systems},'' {\em {IEEE} J. Sel. Topics Signal Process.}, vol.~10,
  pp.~485--500, April 2016.

\bibitem{acceleratedCNN_FPGA_300mW}
B.~{Sun}, L.~{Yang}, P.~{Dong}, W.~{Zhang}, J.~{Dong}, and C.~{Young},
  ``{{Ultra Power-Efficient CNN Domain Specific Accelerator with 9.3TOPS/Watt
  for Mobile and Embedded Applications}},'' {\em arXiv e-prints},
  p.~arXiv:1805.00361, Apr 2018.

\end{thebibliography}
\end{document}